# Comparative Analysis of Software Development Methods between Parallel, V-Shaped and Iterative

Suryanto Nugroho
Department of Information System
STMIK Duta Bangsa
Surakarta, Indonesia

Sigit Hadi Waluyo
Department of Information Management
AMIK Taruna Probolinggo
Probolinggo, Indonesia

Luqman Hakim
Department of Informatics
Muhammadiyah University of Malang
Malang, Indonesia

## ABSTRACT
Any organization that will develop software is faced with a difficult choice of choosing the right software development method. Whereas the software development methods used, play a significant role in the overall software development process. Software development methods are needed so that the software development process can be systematic so that it is not only completed within the right time frame but also must have good quality. There are various methods of software development in System Development Lyfe Cycle (SDLC). Each SDLC method provides a general guiding line about different software development and has different characteristics. Each method of software development has its drawbacks and advantages so that the selection of software development methods should be compatible with the capacity of the software developed. This paper will compare three different software development methods: V-Shaped Model, Parallel Development Model, and Iterative Model with the aim of providing an understanding of software developers to choose the right method.

## 1. INTRODUCTION
The development of information technology has made a new era in human life. As a result of the development of informatics that people use software on every social activity can be said software become one part of modern human civilization. Pressman [8] argues that software becomes a tool in decision making in the contemporary business world, besides software, also serves as the basis of business services that have been made in the form of information systems to support business processes, such as transportation systems, medical, telecommunications, industry, and entertainment.

Currently, an organization depends heavily on software because of the development of information technology. Software developed by the organization serves to assist in performing a complex business process becomes easier, faster and efficient, also the results of data processing through the software more accurate than the computer manually.

Software-dependent organizations must develop software, but each agency is faced with the difficulty of determining which software development method to use. Software development processes play a significant role in the overall software development process; the software development method also determines a software that is produced according to a set time and has a good quality.

System Development Life Cycle or SDLC is a methodology used to perform software development. The concept of the SDLC is composed of several stages starting from the planning stage, the analysis phase, the design stage, the implementation stage and up to the system maintenance period. Different forms of software development models that currently exist in building a framework or framework are based on the concept of SDLC. According to Rainer, Turban, and Potter [9], SDLC is a structured framework that contains following processes in which the system is developed.

Choosing the right software development model allows software developers to set development strategies. Each SDLC model has advantages and disadvantages to being considered when we are going to develop a software because of the influence of the model's characteristics. For that, we need to compare the SDLC model before we determine the usage of the SDLC model. In this paper, we will examine the model of SDLC V-Shaped Model, Parallel Model, and Iterative Model.

## 2. BASIC THEORY
### 2.1 Software
According to Laudon and Laudon [6] argue that software is
1. Command that when executed gives function and work as desired.
2. Data structures that allow programs to manipulate information proportionately.
3. Documents that describe the information and usefulness of the program.

Meanwhile, according to Pressman [8] software is an instruction that when executed produce the desired function and results. The software also means data structures that can manipulate information

Of the two definitions above we can conclude that the software is a command that if executed will get results by the wishes of users. The software is more logical than the physical system element. Therefore, according to Sommerville [10], software should have the following characteristics:
1. Maintainability, the Software must be able to meet changing needs.
2. Dependability, Software must be reliable.
3. Efficiency, Software must be efficient in resource usage.
4. Usability, Software must be used as planned.

### 2.2 Software Engineering
In the opinion of Pressman [8], software engineering is the process of making using the principles of engineering expertise to obtain a software that is economical, reliable and can work efficiently. While Sommerville [10] argues that software is a law that deals with the engineering of all aspects of making software from the system specification stage to the stage of system maintenance after going through the system usage stage.

Of the two definitions above software engineering can be interpreted as a principle in the manufacture of software ranging from the system specification stage to the stage of system maintenance to obtain software that is economical, reliable and efficient.

## 2.3 System Development Life Cycle

System Development Life Cycle or SDLC is a method by which software can be systematically developed and will increase the likelihood of completing software projects within the time limit and maintaining the quality of software products as per the standards.

The SDLC development framework provides a sequence of activities for system designers and developers to follow to develop software. This is often regarded as part of the system development life cycle. Each software development process is divided into several sound stages that allow a software development company to manage the work efficiently to build software products from the required functionality within a particular time frame and budget. All software projects through the stages of requirements collection, business analysis, system design, implementation, and quality assurance testing [5]

## 3. DISCUSSION
### 3.1 SDLC Phase

The SDLC development framework provides a sequence of activities for system designers and developers to follow to develop software. There are five stages in software development using SDLC, namely: planning, analysis, design, implementation, and maintenance.

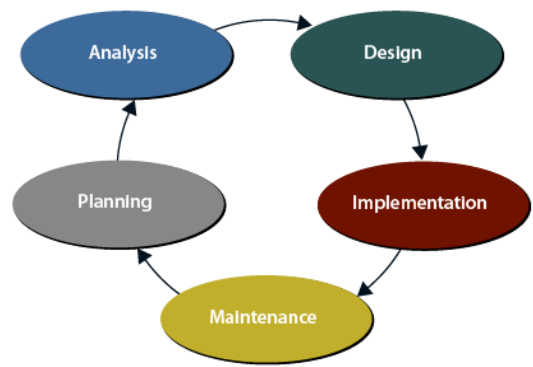

**Figure 1: SDLC Phase**

A. System Planing

The planning stage in software development emphasizes more on the feasibility study aspect. Activities include:
1. Establish and consolidate the development team.
2. Determine the purpose and scope of development.
3. Identify the problems that exist in system development.
4. Detect and evaluate strategies to be used in system development.
5. Determination of technology priority scale and application selection.

B. System Analysis

In the analysis phase, the developer will perform the following activities
1. Conduct literature studies to find a similar case of existing research that can be handled by the system.
2. Brainstorming in the developer team on which case is most appropriately modeled with the system.
3. Classifying problems, solutions, and opportunities that can be solved with the system.
4. Creating analysis and needs of the system.

C. System Design

At this stage, features and operations of the system are described in detail. Activities were undertaken are:
1. Analyze the interaction of objects and functions on the system.
2. Analyze data and create the database schema.
3. Designing user interface.

D. System Implementation

The next stage is the application of the implementation of the design of the previous stages and test. In the implementation, the following activities are carried out:
1. Creation of the database according to the design scheme.
2. Creation of applications based on system design.
3. Testing and repairing the application (debugging).

E. System Maintenance

Performed by the designated admin to keep the system capable of operating properly through the system's ability to adapt itself as needed.

As for the implementation of SDLC, there are various methodologies that can be used. The use of the method will vary depending on the emphasis, whether on business processes or business support data.

## 3.2 Software Development Method

Software development methods are needed so that the software development process can be systematic so that it is not only completed within the right time frame but also must have good quality.

Sommerville [10] argues that the software development method is a representation of the software process. Each development method represents a software development process from a particular point of view.

The software development process model differs from the software development methodology. According to Boehm [3], a method controls through a stage (data specification, allocation of needs, control, etc.), while the model has an interest in providing guidance in a certain order (several steps) in carrying out the tasks that need to be done to complete a project software.

## 3.3 Comparison of Software Development Methods
Comparison of software development methods is done to determine the characteristics of each software development process to facilitate software developers choose the model according to their needs. There are three models compared to this paper: V-shaped model, parallel model, and iterative model.

## 3.4 V-Shaped Model
V-Shaped Model called by Balaji and Murugaiyan [2] "verification and validation model" is a modified version of the Waterfall model. Unlike Waterfall, this one model is not designed linearly. If in the waterfall model the process is run linearly, then in V-model the process is done branched. Software development process in V-Shaped Model depends on verification in the previous stage. In V-Shaped Model gives the relationship between each development stage and Testing stage.

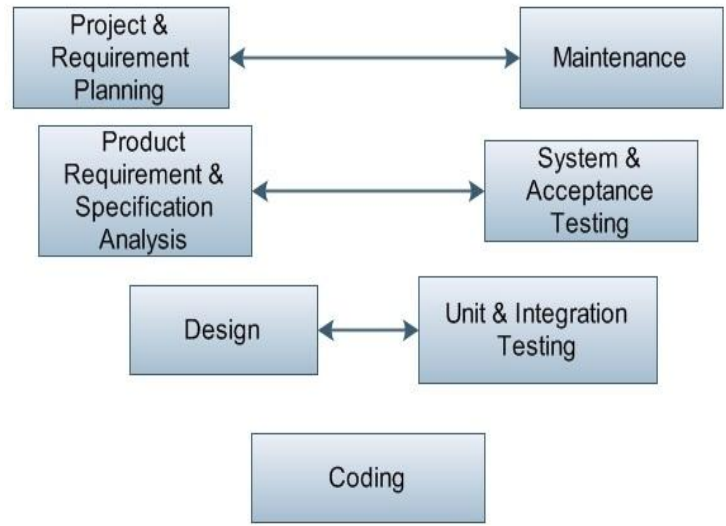

**Figure 2: V-Shaped Model**

Excess of V-Shaped Model
In the opinion of Munassar and Govardhan [7], V Shaped Model has advantages such as:
1. Straightforward and easy to use.
2. Each stage has a certain result.
3. The chance to succeed is higher than the waterfall model because the test is done from the beginning
4. This model is well used for software projects on a small scale

Lacks of V-Shaped Model
While the shortcomings of V-Shaped Model in the opinion of Munassar and Govardhan [7] are:
1. Very rigid as waterfall model.
2. Flexibility is minimal and when will adjust difficult and expensive

3. The software was developed during the implementation phase so that no initial prototype of the software was produced
4. This model does not provide a clear solution to the problems found during the testing phase.

## 3.5 Parallel Model
In the opinion of Ajah and Ugah [1], this model tries to overcome the long time interval between the period of analysis and delivery of the system. A general design for the whole system is done, and then the project is divided into a series of different sub-projects that can be developed and implemented in parallel. This model tries to overcome the weaknesses in the waterfall model so that the concept is made after all subprojects are perfect, then the final integration is done so that delivery is done to the system.

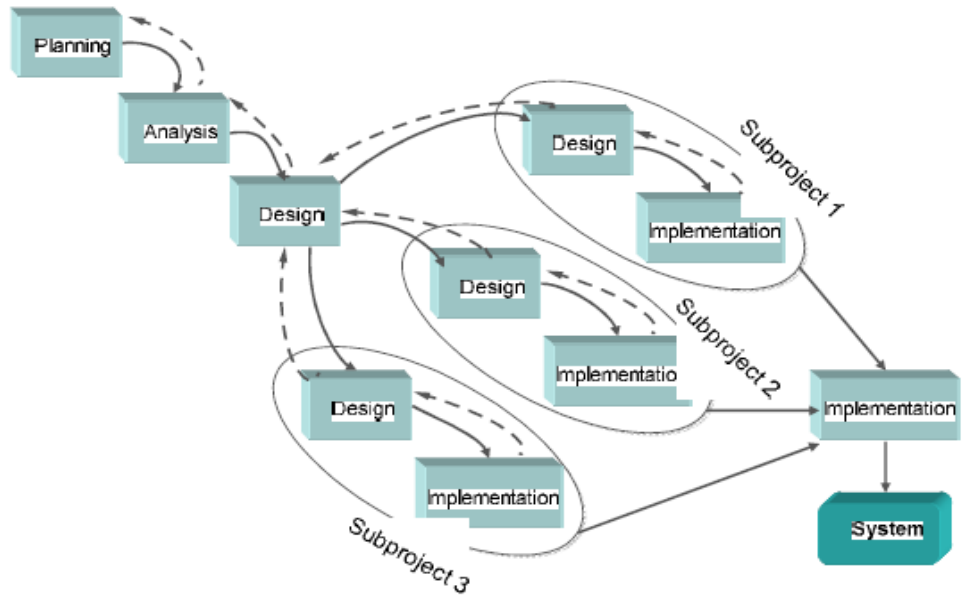

**Figure 3: Parallel Model**

Excess of Parallel Model
Ajah and Ugah [1] argue that the excess parallel model is the time required for system delivery is shorter when compared to the waterfall because some stages have been made in parallel form thus reducing the duration of time.
Lacks of Parallel Model
In the opinion of Ajah and Ugah [1] the parallel shortcomings of the model are:
1. Just like the waterfall design, this model also could not be separated from problems caused by long system delivery.
2. Sometimes subprojects are not independent; this is because the design decision is made in one
3. subproject may affect others.
4. The end of the project may involve significant integration challenges.
5. Not suitable for complex projects and necessary software
While Budi, Siswa, and Abijono [4] added the shortcomings of the parallel model are:

1. System integration has its difficulties. Failure or delay in one subproject impacts the process of integrating the entire system,
2. There is a possibility of difficulty in handling if there is a problem in the subproject simultaneously.

### 3.6 Iterative Model
According to Munassar and Govardhan [7] problems that occur in the waterfall model create demand for software development models that can deliver faster results, require less information, and offer greater flexibility.
With an iterative design, a project is divided into small parts; This allows the development team to show the results of previous processing and gain valuable feedback from users of the system. Often, each iteration is a mini-Waterfall process with feedback from one stage and providing necessary information for the next stage.

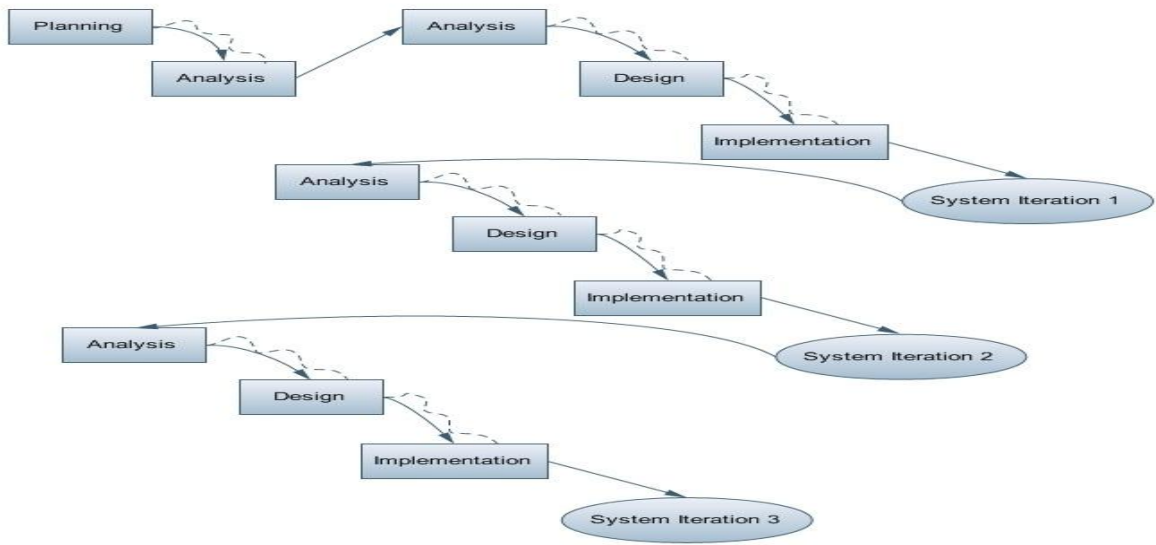

**Figure 4: Iterative Model**

Excess of Iterative Model

In the opinion of Budi et al. [4] the advantages of iterative models are as follows:

1. Feedback can be done continuously from the project owner.
2. Some of the revisions to all specific applications and functions.
3. Work in software development delivered at the beginning of the project

Lacks of Iterative Model

Lacks of iterative modifications include:

1. Although not necessarily a problem for all projects, since minimal initial planning before coding and implementation begins, when using the iteration model, it is possible that unexpected problems in the design or system architecture will arise within the project. Completing this could have potentially disastrous effects on the overall project time and cost of the project, which require many future iterations just to solve one problem.
2. Unlike the waterfall model, which emphasizes almost all user involvement in the early stages of the project, repetitive models often require user engagement throughout the entire process, as each new iteration will likely require testing and feedback from users to evaluate the necessary changes.

## 4. CONCLUSION

From comparison to 3 (three) method above, can be concluded as follows:

1. V-Shaped Model is easy to use but rigid as waterfall model, But the chance of software development success using shaped model is bigger when compared with waterfall model.
2. The time used by the parallel model in the system delivery is shorter than the waterfall model
3. The iterative design requires assistance from users for all stages of system development.
4. In v-shaped models and iterative models, the cost of system development is more expensive than the waterfall model.
5. The project owner's iterative model can perform continuous feedback.

It can be concluded that the success of software development depends on the overall management of software projects. The software development methodology component consists of models, tools (Tools), and procedures. No model really suits all types of organizations, so it takes a further approach to choosing which model is most appropriate to apply to a particular group.

## 5. ACKNOWLEDGMENTS

We express gratitude to God Almighty, Parents, Family, PJJ Aptikom Batch 5 and friends who helped and support so that this work can be completed.